%% file: paper.tex
\documentclass[10pt,conference]{IEEEtran}
\IEEEoverridecommandlockouts
\usepackage{scalerel}
\usepackage{multirow} 
\usepackage{booktabs}
\usepackage[table]{xcolor}
\usepackage[sort&compress, numbers]{natbib}
\usepackage{amsmath,amssymb,amsfonts}
\usepackage{algorithmic}
\usepackage{graphicx}
\usepackage{subcaption}
\usepackage{float}\usepackage{textcomp}
\usepackage{xcolor}
\usepackage{flushend}
\usepackage{textcomp}
\usepackage{hyperref}
\usepackage{comment}
\usepackage{siunitx}
\usepackage[framemethod=TikZ]{mdframed}

\definecolor{seabornGreen}{HTML}{66c2a5}
\definecolor{seabornOrange}{HTML}{fc8d62}
\definecolor{seabornBlueGray}{HTML}{8da0b1}
\definecolor{seabornPink}{HTML}{e78ac3}

\def\BibTeX{{\rm B\kern-.05em{\sc i\kern-.025em b}\kern-.08em
    T\kern-.1667em\lower.7ex\hbox{E}\kern-.125emX}}
\begin{document}


\title{Should Code Models Learn Pedagogically? \\A Preliminary Evaluation of Curriculum Learning for Real-World Software Engineering Tasks}

 \author{\IEEEauthorblockN{Kyi Shin Khant}
 \IEEEauthorblockA{
 \textit{The University of Melbourne}\\
 kyishink@student.unimelb.edu.au}
  \and
 \IEEEauthorblockN{Hong Yi Lin}
 \IEEEauthorblockA{
 \textit{The University of Melbourne}\\
 holin2@student.unimelb.edu.au}
 \and
 \IEEEauthorblockN{Patanamon Thongtanunam}
 \IEEEauthorblockA{
 \textit{The University of Melbourne}\\
 patanamon.t@unimelb.edu.au}
 }

\maketitle

\begin{abstract}
Learning-based techniques, especially advanced
pre-trained models for code have demonstrated capabilities in code understanding and generation,  solving diverse software engineering (SE) tasks.
Despite the promising results, current training approaches may not fully optimize model performance, as they typically involve learning from randomly shuffled training data.
Recent work shows that Curriculum Learning (CL) can improve performance on code-related tasks through incremental learning based on the difficulty of synthetic code.
Yet, the effectiveness of CL with conventional difficulty measures in SE tasks remains largely unexplored.
In this study, we explore two conventional code metrics: code length and cyclomatic complexity to determine the difficulty levels.
We investigate how the pre-trained code model (CodeT5) learns under CL, through the tasks of code clone detection and code summarization.
Our empirical study on the CodeXGLUE benchmark showed contrasting results to prior studies, where the model exhibited signs of catastrophic forgetting and shortcut learning.
Surprisingly, model performance saturates after only the first quartile of training, potentially indicating a limit in the  model's representation capacity and/or the task's inherent difficulty.
Future work should further explore various CL strategies with different code models across a wider range of SE tasks for a more holistic understanding.

\end{abstract}


\section{Introduction}
Learning-based techniques, especially advanced pre-trained models for code have shown potential in code understanding and generation.
These capabilities have proven to be useful for a diverse range of software engineering (SE) tasks, such as vulnerability detection~\cite{linevul}, bug-fixing~\cite{bugfixwild}, and code translation~\cite{codexglue}.
Whilst these results are promising, code models often still struggle with longer~\cite{codestruct} and more complex programs~\cite{austin2021program}.

In recent years, a plethora of research has investigated code-oriented pre-training~\cite{zengslr} as the primary method for improving code models, however, the effects of employing different fine-tuning strategies for downstream SE tasks still lacks attention. 
By comparison, \textit{curriculum learning} (CL), a difficulty-based training strategy inspired by pedagogical learning principles~\cite{clbengio} has demonstrated promising results in domains such as natural language processing and computer vision~\cite{curricsurvey}.
For code understanding, CL has shown potential in synthetic programming challenges involving code completion~\cite{curriccode} and code execution~\cite{liu-etal-2023-code}.
For real-world SE, CL has been effective in tasks such as code clone detection~\cite{wangcl} and automated program repair~\cite{apr_cl}.
However, these studies rely on semantic preserving transformations to create difficult code, rather than natural code difficulty. 
They also do not examine how the model is gaining competency at each stage of CL.

In this study, we set out to investigate whether CL with conventional difficulty measures can enhance the understanding and generative capabilities of pre-trained code models with real-world software.
We also investigate the stages of this learning process, and how each difficulty affects the model.
We experimented with two well-established SE tasks, i.e., code clone detection and code summarization~\cite{zengslr}.
We structured the CL training schedules by organizing the fine-tuning data into four difficulty levels (easy, medium, hard, and very hard) based on conventional difficulty metrics i.e., length ($L$) and cyclomatic complexity ($C$).
We explore both sequential ($s$) and reverse sequential ($r$) CL schedules, i.e., fine-tuning the model from easy to hard, and vice versa.
Through an empirical study of CodeT5~\cite{codet5}, using the CodeXGLUE benchmark~\cite{codexglue}, we address the following two research questions:

\begin{enumerate}
    \item[\textbf{RQ1.}] \textbf{Can curriculum learning with conventional difficulty metrics enhance the performance of code models in code clone detection and code summarization}?\\
\underline{Findings:} In our experiments with CodeT5, CL with conventional difficulty measures did not outperform the traditional random schedule in the two selected tasks.
    \item[\textbf{RQ2.}] \textbf{How does each difficulty level impact model performance during CL?}\\
    \underline{Findings:} Interestingly, model performances reach saturation with only 25\% of the training data. In other words, after initially fine-tuning on either the easy or very hard subset, additional data from the subsequent difficulty levels yield only minimal performance gains.
\end{enumerate}

Our RQ1 findings suggest that, despite promising results of CL with difficult synthetic code, scheduled training based on conventional difficulty of code had little impact on CodeT5's capabilities in the selected tasks.
Instead, RQ2 reveals that CodeT5 can sufficiently learn common patterns for these tasks from limited data, and additional training on other difficulty levels only marginally enhances these capabilities.
Nonetheless, this may be due to the model's limited representation capacity and/or the natural difficulty of the task.
For a more comprehensive evaluation, future studies should include more diverse code models, datasets and difficulty measures.

\textbf{Contribution.}
To the best of our knowledge, we are the first to investigate the learning process of code models during CL with conventional difficulty metrics.
Our study unveils an interesting phenomenon,  providing motivation for further research into how code models learn downstream SE tasks.

\section{Background and Related Work}

\subsection{Neural Code Models}
The success of neural language models in the natural language domain has inspired researchers to apply similar methods to programming languages. 
To alleviate the software engineering workload, a large plethora of research has explored automating SE tasks, such as vulnerability detection~\cite{linevul}, bug-fixing~\cite{bugfixwild}, code translation~\cite{codexglue}, code clone detection~\cite{mou2016} and code summarization~\cite{iyer2016summarizing}. 
Specifically, the techniques involve training neural language models on data from public repositories such as GitHub and Stack Overflow~\cite{codet5}.
Leveraging the naturalness properties of human written code, these neural code models have learned useful patterns that unlock their potential as practical software development assistants~\cite{assetorliability}.

\subsection{Curriculum Learning}

Replicating the pedagogical learning process of humans, CL is a training strategy that gradually increases the difficulty of examples presented to the model~\cite{clbengio}, which contrasts with the conventional unordered training schedule.
CL has been effective for a wide range of model architectures and tasks~\cite{curricnn,curricsurvey}. 
It improves model performance on complex examples by aligning training schedules with the model's skill-acquiring pace~\cite{curricnmt}.
Using generated programs, Naïr et al.~\cite{curriccode} investigated the effects of CL in code completion, whilst Liu et al.~\cite{liu-etal-2023-code} focused on the task of code execution in competitive programming.
Although conventional difficulty measures such as length and cyclomatic complexity were investigated, these studies focused on synthetic programming problems, which may not generalise to real-world software.
In terms of real-world SE, CL has been employed in tasks, such as code clone detection, code search~\cite{wangcl} and automated program repair~\cite{apr_cl}, however, the methods largely focused on augmenting code to create difficult synthetic examples.
Additionally, past studies mainly focused on the final result, rather than investigating how the code model improves during CL, leaving us with an opaque understanding of the process.
Past studies have also yet to investigate the code-to-text task of code summarization.

\section{Study Design}

\subsection{Overview}
To investigate whether CL can improve models' code understanding, we explore two learning schedules: sequential ($s$) and reverse sequential ($r$). 
We organize the training data into four subsets based on difficulty levels and incrementally fine-tune the code models by each subset.
Finally, we compare the performance of models trained incrementally by difficulty level to those trained with the conventional approach i.e., unordered schedule.
We experiment with two well-established SE tasks: code clone detection and code summarization. 
Below, we describe the data preparation, CL schedules, and experimental setup.

\subsection{Data Preparation}
To evaluate CL for real-world SE tasks, we use the CodeXGLUE benchmark~\cite{codexglue}, which includes code collected from GitHub repositories. It has been used as a comprehensive benchmark for evaluating model capabilities across a wide range of real-world SE tasks~\cite{zengslr}.

\textbf{Difficulty Measures:} 
To organize the data into difficulty subsets, we explore two code metrics that determine the difficulty of the code: length ($L$) and cyclomatic complexity ($C$)~\cite{complexitymeasure}.
We measure $L$ by counting tokens separated by spaces. 
This approach aligns with human intuition, as longer texts are often perceived as more challenging to read, understand, and maintain due to a greater number of variables and logical operations~\cite{codenaturalness}.
Prior work has also shown that language models tend to underperform with longer inputs~\cite{long_context}.
We measure $C$, using Lizard, a cyclomatic complexity analyzer,\footnote{\url{https://github.com/terryyin/lizard}} which counts linearly independent paths determined by decision points e.g., if, else, while, for. 
More decision points indicate higher complexity, requiring the model to process multiple execution paths \cite{cyccritiq, metricthresholds}.
Prior work in code clone detection has also shown that language models tend to struggle with examples that are more complex~\cite{llm_ccd}.
For code clone detection, the difficulty score is summed across the program pair.
For code summarization, the difficulty score is determined by the single target program.

\textbf{Difficulty Subsets.} 
We divide the dataset into four difficulty levels based on naturally distributed quartiles. 
The first quartile represents the easiest examples, whilst the fourth quartile contains the most challenging.
Table~\ref{table:subset_data} shows the number of instances in each difficulty level. 
Since subsets vary in size which may influence the model performance, we undersample the larger subsets to match the smallest subset to control the training size of each subset.

\input{tables/subsets}

\subsection{Curriculum Learning Schedules} \label{sec:CL}
For the sake of completeness, we explore two CL schedules i.e., sequential ($s$) and reverse sequential ($r$).
For sequential schedule $s$, we incrementally fine-tune the code model with each subset starting from easy to very hard, resembling the natural pedagogical process.
After fine-tuning each subset, we select the best model checkpoint based on the validation set to be further trained with the next difficulty subset.
For reverse schedule $r$, we incrementally fine-tune from very hard to easy, which explores the effects of learning in an unintuitive manner that does not resemble the natural pedagogical process.
After training on each difficulty subset, we assess the model on the entire test set to capture the model's learning process.

\subsection{Experiment Setup}
To better quantify the impact of CL on initial learning of the selected software engineering tasks, our experiment required a relatively primitive model that has not been subject to vast amounts of training on software engineering datasets.
To this end, we select the 220M parameter CodeT5-base model~\cite{codet5}.
The model was fine-tuned with the original hyper-parameters i.e. learning rate of 5e-5, beam search width of 10, and 1 epoch for code clone detection and 15 epochs for code summarization.
To speed up training, we increased the batch size from 10 to 16 for code clone detection, while maintaining a batch size of 48 for code summarization. 
We used one 32-core server with an NVIDIA A100-80GB GPU for training.

\subsection{Evaluation Metrics}
For code clone detection, we evaluate model performance with F1 score, precision, and recall. 
For code summarization, we evaluate model performance with both the BLEU~\cite{bleu} score and the embedding-based BERTScore~\cite{bertscore}.
Unlike BLEU, BERTScore accounts for model outputs that are textually divergent, yet semantically similar to the ground truth.

\begin{table*}[htbp]
\caption{Stratified Code Clone Detection Results (Length)}
\label{ref:codeclonettable1}
\begin{center}
\resizebox{\textwidth}{!}{%
\begin{tabular}{|l|c|c|c|c|c|c|c|c|c|c|c|c|c|c|c|}
\hline
& \multicolumn{3}{c|}{\textbf{Test - Easy}} & \multicolumn{3}{c|}{\textbf{Test - Medium}} & \multicolumn{3}{c|}{\textbf{Test - Hard}} & \multicolumn{3}{c|}{\textbf{Test - Very Hard}} & \multicolumn{3}{c|}{\textbf{Test - Total}} \\
\cline{2-16}
 & \textbf{F1} & \textbf{Prec} & \textbf{Rec} & \textbf{F1} & \textbf{Prec} & \textbf{Rec} & \textbf{F1} & \textbf{Prec} & \textbf{Rec} & \textbf{F1} & \textbf{Prec} & \textbf{Rec} & \textbf{F1} & \textbf{Prec} & \textbf{Rec} \\
\hline
E & 98.43 & 98.32 & 98.53 & 95.59 & 93.62 & 97.65 & 91.26 & 88.21 & 94.52 & 81.34 & 88.81 & 75.03 & 91.82  & 92.53 & 91.13 \\

E+M & 98.34  & 98.16 & 98.53 & 95.74 & 93.82 & 97.73 & 89.25 & 84.72 & 94.29 & \underline{77.42} & 79.54 & 75.40 & 90.23 & 89.28 & 91.20 \\

E+M+H &  98.04 & 98.07 & \underline{98.07} & 95.70 & 94.57 & 96.86 & 90.49 & 90.30 & 90.68 & 81.99 & 86.04 & 78.31 & 91.65 & 92.45 & \underline{90.87} \\

E+M+H+VH & 98.19 & \underline{97.32} & 99.08 & 95.61 & 94.05 & 97.22 & 92.67 & 89.96 & 95.54 & 88.68 & 88.67 & 88.70 & 93.82 & 92.60 & \textbf{95.06} \\

\hline
VH & 98.06 & 97.37 & 98.79 & 96.46 & 95.16 & 97.81 & 93.89 & 91.76 & \textbf{96.13} & 89.11 & 90.29 & 87.97 & 94.39 & 93.73 & \textbf{95.06} \\

VH+H & \underline{98.02} & 98.07 & \underline{98.07} & 95.70 & 94.57 & 96.86 & 90.49 & 90.30 & 90.68 & 81.99 & 86.04 & 78.31 & 91.65 & 92.45 & \underline{90.87} \\

VH+H+M & 98.84 & 98.46 & \textbf{99.23} & 96.23 & 94.22 & \textbf{98.32} & 91.96 & 88.67 & 95.50 & 79.43 & 84.67 & \underline{74.79} & 91.72 & 91.81 & 91.63 \\

VH+H+M+E & 98.05 & 97.37 & 98.74 & \underline{93.85} & \underline{90.65} & 97.28 & \underline{87.23} & \underline{80.11} & 95.75 & 80.31 & \underline{77.75} & 83.04 & \underline{89.82} & \underline{86.42} & 93.51 \\
\hline

Conventional & \textbf{98.70} & \textbf{99.48} & 97.92 & \textbf{96.47} & \textbf{97.77} & \underline{95.21} & \textbf{95.11} & \textbf{96.12} & \underline{94.12} & \textbf{90.75} & \textbf{91.77} & \textbf{89.75} & \textbf{95.24} & \textbf{96.26} & 94.23 \\
\hline
\multicolumn{13}{l}{\footnotesize (E)Easy, (M)Medium, (H)Hard, (VH)Very Hard, \textbf{Best Score}, \underline{Worst Score}, \textbf{Prec}ision, \textbf{Rec}all}
\end{tabular}%
}
\label{t8}
\end{center}
\end{table*}

\begin{table*}[htbp]
\caption{Stratified Code Clone Detection Results (Cyclomatic Complexity)}
\label{ref:codeclonettable2}
\begin{center}
\resizebox{\textwidth}{!}{%
\begin{tabular}{|l|c|c|c|c|c|c|c|c|c|c|c|c|c|c|c|}
\hline
& \multicolumn{3}{c|}{\textbf{Test - Easy}} & \multicolumn{3}{c|}{\textbf{Test - Medium}} & \multicolumn{3}{c|}{\textbf{Test - Hard}} & \multicolumn{3}{c|}{\textbf{Test - Very Hard}} & \multicolumn{3}{c|}{\textbf{Test - Total}} \\
\cline{2-16}
 & \textbf{F1} & \textbf{Prec} & \textbf{Rec} & \textbf{F1} & \textbf{Prec} & \textbf{Rec} & \textbf{F1} & \textbf{Prec} & \textbf{Rec} & \textbf{F1} & \textbf{Prec} & \textbf{Rec} & \textbf{F1} & \textbf{Prec} & \textbf{Rec} \\
\hline
E & 98.61 & 98.47 & 98.75 & 95.66 & 95.01 & 96.32 & 91.66 & 89.76 & 93.65 & 87.88 & 89.09 & 86.70 & 93.40 & 93.11 & 93.70 \\

E+M & \underline{96.16}  & \underline{94.01} & 98.42 & 95.71 & 93.33 & \textbf{98.21} & 91.53 & 87.33 & \textbf{96.15} & 87.19 & 87.51 & 86.86 & 93.15 & 91.39 & 94.97 \\

E+M+H &  98.42 & 98.08 & 98.77 & 96.03 & 96.44 & \underline{95.61} & 91.35 & 91.54 & \underline{91.16} & 88.56 & 88.15 & 88.97 & 93.46 & 93.43 & 93.49 \\

E+M+H+VH & 97.29 & 95.71 & 98.93 & 94.41 & \underline{91.57} & 97.44 & 90.76 & 86.32 & 95.68 & 87.72 & 85.17 & 90.42 & 92.55 & 89.75 & \textbf{95.53} \\

\hline
VH & \textbf{98.63} & 98.44 & 98.81 & 95.70 & 94.85 & 96.56 & 92.21 & 90.34 & 94.15 & \textbf{90.48} & 89.43 & \textbf{91.56} & 94.31 & 93.34 & 95.29 \\

VH+H & 98.42 & 98.08 & 98.77 & 96.03 & 96.44 & \underline{95.61} & 91.35 & 91.54 & \underline{91.16} & 88.56 & 88.15 & 88.97 & 93.62 & 93.54 & 93.70 \\

VH+H+M & 98.52 & 97.96 & \textbf{99.09} & 95.69 & 93.57 & 97.91 & 92.05 & 88.68 & 95.69 & 86.87 & 87.15 & 86.60 & 93.27 & 91.96 & 94.62 \\

VH+H+M+E & 97.30 & 96.08 & 98.60 & \underline{94.33} & 92.54 & 96.18 & \underline{87.81} & \underline{83.58} & 92.51 & \underline{81.71} & \underline{78.88} & \underline{84.75} & \underline{90.19} & \underline{87.65} & \underline{92.88} \\
\hline

Conventional & 98.58 & \textbf{99.20} & \underline{97.96} & \textbf{96.52} & \textbf{97.16} & 95.88 & \textbf{93.12} & \textbf{93.47} & 92.76 & 89.69 & \textbf{89.83} & 89.55 & \textbf{94.46} & \textbf{94.89} & 94.04 \\
\hline
\multicolumn{13}{l}{\footnotesize (E)Easy, (M)Medium, (H)Hard, (VH)Very Hard, \textbf{Best Score}, \underline{Worst Score}, \textbf{Prec}ision, \textbf{Rec}all}
\end{tabular}%
}
\label{t8}
\end{center}
\end{table*}

\begin{table*}[htbp]
\caption{Stratified Code Summarization Results (Length)}
\label{ref:codesummarisationttable1}
\begin{center}
\resizebox{\textwidth}{!}{
\begin{tabular}{|l|c|c|c|c|c|c|c|c|c|c|}
\hline
& \multicolumn{2}{c|}{\textbf{Test - Easy}} & \multicolumn{2}{c|}{\textbf{Test - Medium}} & \multicolumn{2}{c|}{\textbf{Test - Hard}} & \multicolumn{2}{c|}{\textbf{Test - Very Hard}} & \multicolumn{2}{c|}{\textbf{Test - Total}}\\

\cline{2-11} 
 & \textbf{BLEU} & $\textbf{BS}_{F1}$ & \textbf{BLEU} & $\textbf{BS}_{F1}$ & \textbf{BLEU} & $\textbf{BS}_{F1}$  & \textbf{BLEU} & $\textbf{BS}_{F1}$ & \textbf{BLEU} & $\textbf{BS}_{F1}$ \\
 
\hline
E & 24.34 & 91.39 & 21.64 & 91.50 & \underline{18.44} & \underline{91.16} & \underline{16.50} & 90.64  & 20.23 & 91.17 \\

E+M & \textbf{24.77} & 91.53 & 22.46 & 91.65 & 19.14 & 91.30 & 17.23 & 90.77  & 20.90 & 91.31 \\

E+M+H & 24.39  & 91.50 & 22.07 & 91.67 & 18.97 & 91.32 & 17.00 & 90.76 & 20.61 & 91.31 \\

E+M+H+VH & 23.98 & 91.45 & 21.5 & 91.59 & 19.06 & 91.26 & 17.19 & 90.76  & 20.43 & 91.27 \\
\hline

VH & \underline{21.12} & \underline{91.35} & 19.85 & \underline{91.42} & 18.72 & 91.24 & 17.11 & 90.50  & \underline{19.20} & \underline{91.13} \\

VH+H & 22.53 & 91.45 & \underline{19.66}  & 91.43 & 18.85 & 91.28 & 17.07 & \underline{90.47}  & 19.53 & 91.15 \\

VH+H+M & 24.51 & \textbf{91.81} & 22.23 & 91.66 & 19.05 & 91.32 & 17.38 & 90.49  & 20.79 & 91.32 \\

VH+H+M+E & 24.47 & \textbf{91.81} & 22.12 & 91.64 & \textbf{19.26}  & 91.28  & \textbf{17.71}  & 90.48  & \textbf{20.89} & 91.30 \\
\hline

Conventional & 24.36 & 91.53 & \textbf{22.54} & \textbf{91.71} & 18.99 & \textbf{91.34} & 17.22 & \textbf{90.80}  & 20.77 & \textbf{91.35} \\
\hline
\multicolumn{11}{l}{\footnotesize (E)Easy, (M)Medium, (H)Hard, (VH)Very Hard, \textbf{Best Score}, \underline{Worst Score}, $\textbf{B}\text{ERT}\textbf{S}\text{core}_{F1}$: RoBERTa-Large Layer 17}
\end{tabular}
}
\label{t12}
\end{center}
\end{table*}

\begin{table*}[htbp]
\caption{Stratified Code Summarization Results (Cyclomatic Complexity)}
\label{ref:codesummarisationttable2}
\begin{center}
\resizebox{\textwidth}{!}{
\begin{tabular}{|l|c|c|c|c|c|c|c|c|c|c|}
\hline
& \multicolumn{2}{c|}{\textbf{Test - Easy}} & \multicolumn{2}{c|}{\textbf{Test - Medium}} & \multicolumn{2}{c|}{\textbf{Test - Hard}} & \multicolumn{2}{c|}{\textbf{Test - Very Hard}} & \multicolumn{2}{c|}{\textbf{Test - Total}}\\

\cline{2-11} 
 & \textbf{BLEU} & $\textbf{BS}_{F1}$ & \textbf{BLEU} & $\textbf{BS}_{F1}$ & \textbf{BLEU} & $\textbf{BS}_{F1}$  & \textbf{BLEU} & $\textbf{BS}_{F1}$ & \textbf{BLEU} & $\textbf{BS}_{F1}$ \\
 
\hline
E & 23.22 & \underline{91.46} & 21.45 & 91.48 & \underline{18.19} & 90.91 & \underline{16.72} & 90.62  & 19.89 &91.12 \\

E+M & 23.38 & 91.62 & 21.68 & 91.60 & 18.54 & 91.03 & 17.04 & 90.76  & 20.16 & 91.25 \\

E+M+H & \textbf{23.64}  & 91.65 & \textbf{22.13} & \textbf{91.64} & 18.81 & 91.12 & 17.44 & \textbf{90.83} & \textbf{20.51} & 91.31 \\

E+M+H+VH & 23.18 & 91.57 & 21.81 & 91.53 & \textbf{18.83} & 91.08 & 17.29 & 90.78  & 20.28 & 91.24 \\
\hline

VH & \underline{21.45} & 91.48 & \underline{19.63} & \underline{91.34} & 18.21  & \underline{90.89} & 16.89 & \underline{90.47}  & \underline{19.05} & \underline{91.04} \\

VH+H & 21.85 & 91.59 & 19.93  & 91.44 & \textbf{18.83} & 91.11 & 17.33 & 90.59  & 19.49 & 91.18 \\

VH+H+M & 22.06 & 91.66 & 21.63 & 91.58 & 18.67 & 91.07 & 17.32 & 90.53  & 19.92 & 91.21 \\

VH+H+M+E & 22.96 & \textbf{91.75} & 21.80 & 91.58 & 18.73  & 91.07  & 17.49 & 90.53  & 20.24 & 91.23 \\
\hline

Conventional & 23.05 & 91.71 & 22.02 & 91.62 & 18.79 & \textbf{91.13} & \textbf{17.56} & 90.82  & 20.44 & \textbf{91.32} \\
\hline
\multicolumn{11}{l}{\footnotesize (E)Easy, (M)Medium, (H)Hard, (VH)Very Hard, \textbf{Best Score}, \underline{Worst Score}, $\textbf{B}\text{ERT}\textbf{S}\text{core}_{F1}$: RoBERTa-Large Layer 17}
\end{tabular}
}
\label{t12}
\end{center}
\end{table*}

\section{Results}
\label{sec:results}
\subsection{Model Performance with Curriculum Learning (RQ1)}

To answer RQ1, we compare models trained with CL to those trained conventionally based on test set performance.
We evaluate four CL strategies based on a combination of the two code metrics, i.e., length ($L$) and cyclomatic complexity ($C$); and the two CL schedules, i.e., sequential ($s$) and reverse ($r$).
For the conventional method, we mix the four subsets into one unordered training set
to fine-tune the model~\cite{codet5}. 
We report two results for the conventional method, the training size differs when it was undersampled to match $L$ or $C$ subsets.


\textbf{Code Clone Detection.} Evaluating based on the whole test set, the conventional method yields the F1 scores, i.e., 95.24 ($L$) and 94.46 ($C$).
In contrast, CL strategies achieve lower  F1-score, i.e., 93.82 ($L_s$), 89.82 ($L_r$), 92.55 ($C_s$), and 90.19 ($C_r$).
Tables~\ref{ref:codeclonettable1} and~\ref{ref:codeclonettable2} show the results across the difficulty subsets.
For easy and medium test subsets, all methods perform similarly as the difference in F1 scores lies within $\Delta$1.29 ($C - C_{s}$) and $\Delta$2.62 ($L - L_{r}$), respectively.
For hard and very hard examples, the divergence in performance is more pronounced.
We find that $s$ schedules only exhibit minor gaps of at most $\Delta$2.44 (Hard: $L - L_{s}$) in F1 scores to the conventional method, whilst, $r$ schedules exhibit major gaps of up to $\Delta$10.44 (Very Hard: $L - L_{r}$) in F1 scores.
These findings resemble symptoms of catastrophic forgetting, which is strongly correlated with task complexity~\cite{clcontinual}.
The difference in behavior between $s$ and $r$ schedules suggests that harder examples presented at the start of the training schedule may be easily forgotten by the end.
Conversely, the model's understanding of easy examples scheduled at the beginning is still largely intact by the end.

\textbf{Code Summarization.} The conventional method yields the BLEU scores of 20.77 ($L$) and 20.44 ($C$).
In contrast, CL strategies achieve mixed results, but marginally different, i.e., 20.43 ($L_s$), 20.89 ($L_r$), 20.28 ($C_s$), and 20.24 ($C_r$).
Tables~\ref{ref:codesummarisationttable1} and~\ref{ref:codesummarisationttable2} show the results across the difficulty subsets.
We find that all strategies have negligible differences in performance across the difficulties.
The gap in BLEU scores is at most $\Delta$0.38 ($L - L_{s}$) for easy, $\Delta$1.04 ($L - L_{s}$) for medium, $\Delta$0.27 for hard ($L_{r} - L$), and $\Delta$0.49 ($L_{r} - L$) for very hard. 
Similar trends are also reflected in the $\text{BERTScore}_{F1}$ results.
These findings suggest that model learning is invariant to the difficulty schedule.
Compared to the prior work~\cite{curriccode}, the synthetic programs in their experiment were deliberately constructed without descriptive names, e.g., using variable names \texttt{a}, \texttt{b}, \texttt{c}. 
Unlike the prior work, our experiment uses real-world programs that have meaningful names for variables, functions, and other semantic cues designed by human developers.
It is possible that models may depend more on these cues rather than program functionality.
Such a behavior is canonically known as shortcut learning~\cite{shortcutlearning}.

In summary, we find that CL schedules with conventional difficulty measures are not conducive to improving CodeT5's performance in both tasks. 
For code clone detection, the models seem to exhibit catastrophic forgetting, which significantly affects the model's competence on harder programs.
For code summarization, the models show invariance to the different schedules, which is possibly due to relaxation on the task's demand for understanding program semantics.

\begin{figure}[t]
\centerline{\includegraphics[width=\columnwidth]{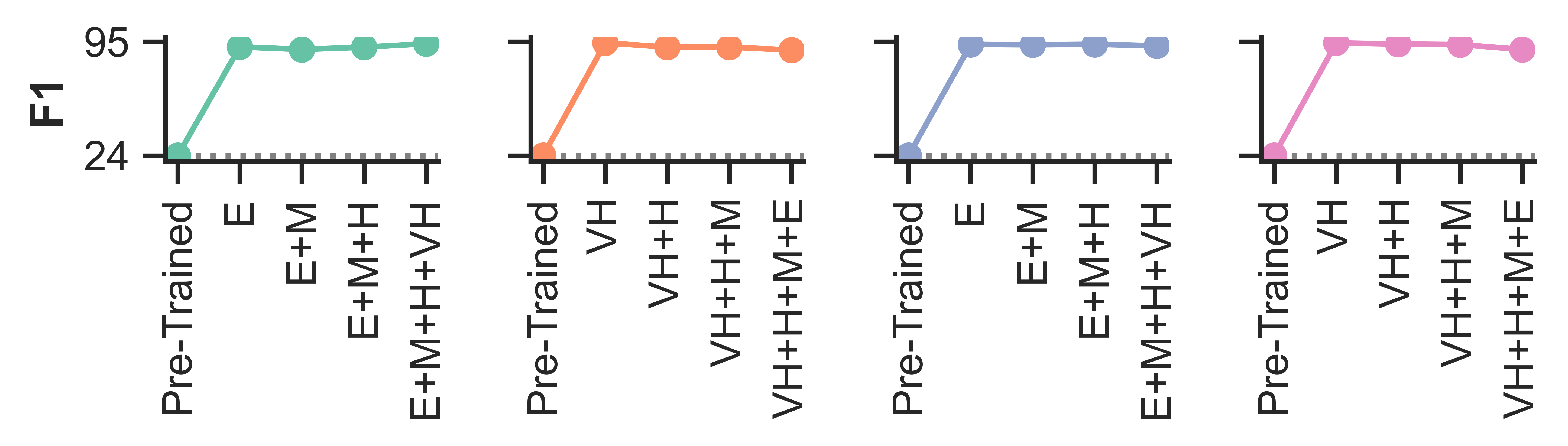}}
\caption{Code Clone Detection (\textcolor{seabornGreen}{$L_{s}$}, \textcolor{seabornOrange}{$L_{r}$}, \textcolor{seabornBlueGray}{$C_{s}$}, \textcolor{seabornPink}{$C_{r}$})}
\label{fig:codeclone}
\end{figure}

\begin{figure}[t]
\centerline{\includegraphics[width=\columnwidth]{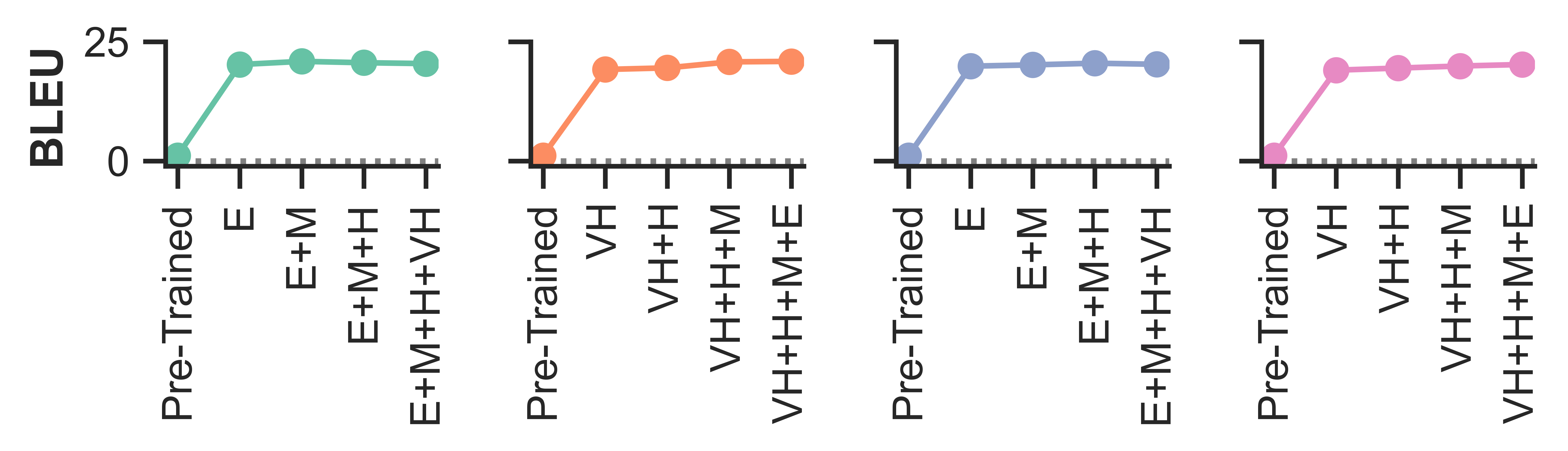}}
\caption{Code Summarization (\textcolor{seabornGreen}{$L_{s}$}, \textcolor{seabornOrange}{$L_{r}$}, \textcolor{seabornBlueGray}{$C_{s}$}, \textcolor{seabornPink}{$C_{r}$})}
\label{fig:codesummarization}
\end{figure}

\subsection{Impact of Each Difficulty Level (RQ2)}
To answer RQ2, we compare the model performance on the entire test set after training on each difficulty subset.
Figures~\ref{fig:codeclone} and~\ref{fig:codesummarization} show the results of the four CL strategies (i.e., $L_{s}$, $L_{r}$, $C_{s}$, $C_{r}$) for code clone detection and code summarization.

\textbf{Code Clone Detection.}
Figure~\ref{fig:codeclone}  shows that high accuracy is already achieved by the model after training on the first quartile for all four strategies. 
Table~\ref{ref:codeclonettable1} shows that for the $L_{s}$ schedule, the model trained only on the easy subset could already achieve a comparable F1 score to the model that has completed the full conventional training, where the difference in F1 scores is $\Delta$0.27 for the easy subset, $\Delta$0.88 for the medium subset and $\Delta$3.85 for the hard subset.
This early model state only struggled against the fully trained model for the very hard subset, with a $\Delta$9.41 drop in F1 score.
Table~\ref{ref:codeclonettable2} shows that for the $C_{s}$ schedule, the model trained on the easy subset was consistently comparable to the fully trained model for all subsets, where the gap is at most $\Delta$1.81 (Very Hard) F1 score.
This demonstrates that learning from easy examples in terms of $C$, already equipped the model with sufficient knowledge to generalise to very hard examples.
For $L_{r}$, we find that the gap across the difficulties between the model trained only on the very hard subset and the fully trained model is at most $\Delta$1.64 (Very Hard) F1 score.
For $C_{r}$, we find that this gap is at most $\Delta$0.91 (Hard) F1 score.
This demonstrates that  learning from very hard examples in terms of either $L$ or $C$, is also sufficient to generalise to the rest of the difficulties.

\textbf{Code Summarization.}
Figure~\ref{fig:codesummarization} shows that the model achieves high accuracy after training on only the first quartile for all four strategies. 
Table~\ref{ref:codesummarisationttable1} shows that for the $L_{s}$ schedule, the gap across the difficulties between the model trained on only the easy subset and the fully trained model is at most $\Delta$0.72 BLEU (Very Hard).
Table~\ref{ref:codesummarisationttable2} shows that for the $C_{s}$ schedule, this gap is at most $\Delta$0.84 BLEU (Very Hard).
These trends also translate directly to the $r$ schedules.
For the $L_{r}$ schedule, the gap across the difficulties between the model trained on only the very hard subset and the fully trained model is at most $\Delta$3.24 BLEU (Easy).
For the $C_{r}$ schedule, this gap is at most $\Delta$2.39 BLEU (Medium).
Similar to code clone detection, we find that both the easy and very hard subsets were sufficient for the model to learn useful knowledge that could be generalized to the rest of the difficulties.

Interestingly, all CL schedules show saturation in model performance after learning from the first quartile, indicating that minimal gains are extracted from the rest of the training process.
In other words, the model only needs to learn from the easy or very hard subsets to sufficiently generalize to the rest of the test set for both tasks.
These results may suggest a limitation of the model's representation capacity or a ceiling in the task's inherent difficulty.

\section{Threats to Validity} 
Our study focuses on two SE tasks which 
do not generalize to the full spectrum of SE tasks.
Our results are based on the CodeT5, which may not generalize to other models.
For the code summarization task, we focused exclusively on Java, whereas CodeXGLUE offers sub-tasks for several other programming languages. By restricting our analysis to Java, we may have missed important insights that could emerge from cross-language comparisons.
We focus on only two difficulty measures, which do not represent all the potential aspects of difficulty that challenges the model.
We only explored two predefined CL schedules, which may exhibit different behaviour to automatic self-paced schedules~\cite{curricsurvey}.
The cannonical metrics used to measure code summarisation are not perfect reflections of summary quality, thus, there may be discrepancy between evaluation results and actual task competency.

\section{Conclusion \& Future Work} 
In this work, we explore the process of curriculum learning with conventional difficulty measures - a difficulty-based training strategy inspired by pedagogical learning principles - for a neural code model (CodeT5) in solving real-world SE tasks.
Whilst past studies achieved promising results based on code that was synthetically constructed to be difficult, our work shows contrastive findings for CL with conventional difficulty measures in clone detection and code summarization.
Interestingly, our results show symptoms of catastrophic forgetting and shortcut learning, as well as limitions in the the model's representation capacity and/or the task's inherent difficulty.

Whilst this study investigated two difficulty measures in length and cyclomatic complexity, there are a myriad of potential difficulty measures that have yet to be explored e.g., clone type, functionality.
Additional experiments across a more comprehensive set of difficulty measures, with large scale reruns, could reveal more insights.
Future research should also explore other CL strategies, with diverse neural code models for a broader range of SE tasks with multiple programming languages.
This would help determine the effectiveness of CL across a wider variety of scenarios, providing a more comprehensive understanding of its generalization capabilities.

\textbf{Data Availability.}
All the materials produced from this study are available on Zenodo\footnote{\url{https://zenodo.org/records/14059483}}.

\bibliographystyle{IEEEtran}
\bibliography{references}
\end{document}

%% file: tables/subsets.tex
\definecolor{forestgreen}{RGB}{34, 139, 34}  

\begin{table*}
\begin{center}
\caption{Data Sizes Across Four Difficulty Subsets}
\begin{tabular}{|r|l|l|c|c|l|c|c||l|c|c|l|c|c|}
\hline

& & \multicolumn{6}{c||}{\textbf{Code Clone Detection}}
  & \multicolumn{6}{c|}{\textbf{Code Summarization}}\\

\cline{2-14}
& \textbf{Difficulty} & \multicolumn{3}{c|}{\textbf{Length}}
 &  \multicolumn{3}{c||}{\textbf{Cyclomatic Complexity}}
  & \multicolumn{3}{c|}{\textbf{Length}}
 &  \multicolumn{3}{c|}{\textbf{Cyclomatic Complexity}}
 
 \\ \hline
& \textbf{Subset} & \textbf{Train} & \textbf{Val.} & \textbf{Test}  & \textbf{Train} & \textbf{Val.} & \textbf{Test} 
     & \textbf{Train} & \textbf{Val.} & \textbf{Test}  & \textbf{Train} & \textbf{Val.} & \textbf{Test} \\
\hline
\parbox[t]{2mm}{\multirow{4}{*}{\rotatebox[origin=c]{90}{\textbf{Original}}}}
&\textbf{Easy} & 224K & 86K & 109K 
             & 219K & 84K & 105K 
            & 40K & 1.4K & 2.3K 
            & 46K & 1.7K & 2.6K\\ 
&\textbf{Medium}  & 224K & 106K  & 110K  
             & 210K & 97K & 107K 
             & 40K & 1.3K & 2.7K 
            & 36K & 1.2K & 2.6K \\ 
& \textbf{Hard}  & 225K  & 112K  & 104K  
              & 240K & 121K & 116K 
             & 42K & 1.3K & 2.9K 
            & 43K & 1.3K & 3.2K \\ 
& \textbf{Very Hard}  & 226K  & 110K  & 90K 
             & 229K & 111K & 86K 
             & 41K  & 1K & 2.9K 
             & 37K & 933 & 2.5K \\ 
\hline
\multicolumn{2}{|l|}{\textbf{Balanced$^\dag$ }} & 224K & 86K & 90K 
             & 210K & 84K & 86K 
            & 40K & 1K & 2.3K 
            & 36K & 933 & 2.5K\\ 
\hline
\multicolumn{14}{l}{\footnotesize \dag The subsets were undersampled to achieve equal sizes with the smallest subset.}

\end{tabular}%
\label{table:subset_data}
\end{center}
\end{table*}

%% file: paper.bbl
\begin{thebibliography}{10}
\providecommand{\url}[1]{#1}
\csname url@samestyle\endcsname
\providecommand{\newblock}{\relax}
\providecommand{\bibinfo}[2]{#2}
\providecommand{\BIBentrySTDinterwordspacing}{\spaceskip=0pt\relax}
\providecommand{\BIBentryALTinterwordstretchfactor}{4}
\providecommand{\BIBentryALTinterwordspacing}{\spaceskip=\fontdimen2\font plus
\BIBentryALTinterwordstretchfactor\fontdimen3\font minus \fontdimen4\font\relax}
\providecommand{\BIBforeignlanguage}[2]{{%
\expandafter\ifx\csname l@#1\endcsname\relax
\typeout{** WARNING: IEEEtran.bst: No hyphenation pattern has been}%
\typeout{** loaded for the language `#1'. Using the pattern for}%
\typeout{** the default language instead.}%
\else
\language=\csname l@#1\endcsname
\fi
#2}}
\providecommand{\BIBdecl}{\relax}
\BIBdecl

\bibitem{linevul}
M.~Fu and C.~Tantithamthavorn, ``Linevul: {A} transformer-based line-level vulnerability prediction,'' in \emph{MSR}.\hskip 1em plus 0.5em minus 0.4em\relax New York, NY, USA: ACM, 2022, pp. 608--620.

\bibitem{bugfixwild}
M.~Tufano, C.~Watson, G.~Bavota, M.~D. Penta, M.~White, and D.~Poshyvanyk, ``An empirical study on learning bug-fixing patches in the wild via neural machine translation,'' \emph{TOSEM}, vol.~28, no.~4, pp. 19:1--19:29, 2019.

\bibitem{codexglue}
S.~Lu, D.~Guo, S.~Ren, J.~Huang, A.~Svyatkovskiy, A.~Blanco, C.~B. Clement, D.~Drain, D.~Jiang, D.~Tang, G.~Li, L.~Zhou, L.~Shou, L.~Zhou, M.~Tufano, M.~Gong, M.~Zhou, N.~Duan, N.~Sundaresan, S.~K. Deng, S.~Fu, and S.~Liu, ``Codexglue: {A} machine learning benchmark dataset for code understanding and generation,'' in \emph{NeurIPS Datasets and Benchmarks}, 2021.

\bibitem{codestruct}
H.~Y. Lin and P.~Thongtanunam, ``Towards automated code reviews: Does learning code structure help?'' in \emph{SANER}.\hskip 1em plus 0.5em minus 0.4em\relax NJ, USA: IEEE, 2023, pp. 703--707.

\bibitem{austin2021program}
J.~Austin, A.~Odena, M.~I. Nye, M.~Bosma, H.~Michalewski, D.~Dohan, E.~Jiang, C.~J. Cai, M.~Terry, Q.~V. Le, and C.~Sutton, ``Program synthesis with large language models,'' \emph{CoRR}, vol. abs/2108.07732, 2021.

\bibitem{zengslr}
Z.~Zeng, H.~Tan, H.~Zhang, J.~Li, Y.~Zhang, and L.~Zhang, ``An extensive study on pre-trained models for program understanding and generation,'' in \emph{ISSTA}.\hskip 1em plus 0.5em minus 0.4em\relax New York, NY, USA: ACM, 2022, pp. 39--51.

\bibitem{clbengio}
Y.~Bengio, J.~Louradour, R.~Collobert, and J.~Weston, ``Curriculum learning,'' in \emph{ICML}, vol. 382.\hskip 1em plus 0.5em minus 0.4em\relax New York, NY, USA: ACM, 2009, pp. 41--48.

\bibitem{curricsurvey}
X.~Wang, Y.~Chen, and W.~Zhu, ``A survey on curriculum learning,'' \emph{TPAMI}, vol.~44, no.~9, pp. 4555--4576, 2022.

\bibitem{curriccode}
M.~Na{\"{\i}}r, K.~M. Yamani, L.~S. L'Hadj, and R.~Baghdadi, ``Curriculum learning for small code language models,'' in \emph{ACL - Student Research Workshop}, 2024, pp. 531--542.

\bibitem{liu-etal-2023-code}
C.~Liu, S.~Lu, W.~Chen, D.~Jiang, A.~Svyatkovskiy, S.~Fu, N.~Sundaresan, and N.~Duan, ``Code execution with pre-trained language models,'' in \emph{ACL}, 2023, pp. 4984--4999.

\bibitem{wangcl}
D.~Wang, Z.~Jia, S.~Li, Y.~Yu, Y.~Xiong, W.~Dong, and X.~Liao, ``Bridging pre-trained models and downstream tasks for source code understanding,'' in \emph{ICSE}.\hskip 1em plus 0.5em minus 0.4em\relax New York, NY, USA: ACM, 2022, pp. 287--298.

\bibitem{apr_cl}
S.~Hao, X.~Shi, H.~Liu, and Y.~Shu, ``Enhancing code language models for program repair by curricular fine-tuning framework,'' in \emph{ICSME}.\hskip 1em plus 0.5em minus 0.4em\relax NJ, USA: IEEE, 2023, pp. 136--146.

\bibitem{codet5}
Y.~Wang, W.~Wang, S.~R. Joty, and S.~C.~H. Hoi, ``Codet5: Identifier-aware unified pre-trained encoder-decoder models for code understanding and generation,'' in \emph{EMNLP}.\hskip 1em plus 0.5em minus 0.4em\relax ACL, 2021, pp. 8696--8708.

\bibitem{mou2016}
L.~Mou, G.~Li, L.~Zhang, T.~Wang, and Z.~Jin, ``Convolutional neural networks over tree structures for programming language processing,'' in \emph{AAAI}.\hskip 1em plus 0.5em minus 0.4em\relax {AAAI} Press, 2016, pp. 1287--1293.

\bibitem{iyer2016summarizing}
S.~Iyer, I.~Konstas, A.~Cheung, and L.~Zettlemoyer, ``Summarizing source code using a neural attention model,'' in \emph{ACL}, 2016.

\bibitem{assetorliability}
A.~M. Dakhel, V.~Majdinasab, A.~Nikanjam, F.~Khomh, M.~C. Desmarais, and Z.~M.~J. Jiang, ``Github copilot {AI} pair programmer: Asset or liability?'' \emph{J. Syst. Softw.}, vol. 203, p. 111734, 2023.

\bibitem{curricnn}
A.~Graves, M.~G. Bellemare, J.~Menick, R.~Munos, and K.~Kavukcuoglu, ``Automated curriculum learning for neural networks,'' in \emph{ICML}, vol.~70.\hskip 1em plus 0.5em minus 0.4em\relax PMLR, 2017, pp. 1311--1320.

\bibitem{curricnmt}
E.~A. Platanios, O.~Stretcu, G.~Neubig, B.~P{\'{o}}czos, and T.~M. Mitchell, ``Competence-based curriculum learning for neural machine translation,'' in \emph{NAACL-HLT}.\hskip 1em plus 0.5em minus 0.4em\relax ACL, 2019, pp. 1162--1172.

\bibitem{complexitymeasure}
T.~J. McCabe, ``A complexity measure,'' \emph{TSE.}, vol.~2, no.~4, pp. 308--320, 1976.

\bibitem{codenaturalness}
M.~Allamanis, E.~T. Barr, P.~T. Devanbu, and C.~Sutton, ``A survey of machine learning for big code and naturalness,'' \emph{CSUR}, vol.~51, no.~4, pp. 81:1--81:37, 2018.

\bibitem{long_context}
N.~F. Liu, K.~Lin, J.~Hewitt, A.~Paranjape, M.~Bevilacqua, F.~Petroni, and P.~Liang, ``Lost in the middle: How language models use long contexts,'' \emph{TACL}, vol.~12, pp. 157--173, 2024.

\bibitem{cyccritiq}
M.~J. Shepperd, ``A critique of cyclomatic complexity as a software metric,'' \emph{Softw. Eng. J.}, vol.~3, no.~2, pp. 30--36, 1988.

\bibitem{metricthresholds}
T.~L. Alves, C.~Ypma, and J.~Visser, ``Deriving metric thresholds from benchmark data,'' in \emph{ICSM}.\hskip 1em plus 0.5em minus 0.4em\relax NJ, USA: IEEE, 2010, pp. 1--10.

\bibitem{llm_ccd}
M.~Khajezade, J.~J. Wu, F.~H. Fard, G.~Rodr{\'{\i}}guez{-}P{\'{e}}rez, and M.~S. Shehata, ``Investigating the efficacy of large language models for code clone detection,'' in \emph{ICPC}.\hskip 1em plus 0.5em minus 0.4em\relax New York, NY, USA: ACM, 2024, pp. 161--165.

\bibitem{bleu}
K.~Papineni, S.~Roukos, T.~Ward, and W.~Zhu, ``Bleu: a method for automatic evaluation of machine translation,'' in \emph{ACL}, 2002, pp. 311--318.

\bibitem{bertscore}
T.~Zhang, V.~Kishore, F.~Wu, K.~Q. Weinberger, and Y.~Artzi, ``Bertscore: Evaluating text generation with {BERT},'' in \emph{ICLR}, 2020.

\bibitem{clcontinual}
C.~V. Nguyen, A.~Achille, M.~Lam, T.~Hassner, V.~Mahadevan, and S.~Soatto, ``Toward understanding catastrophic forgetting in continual learning,'' \emph{CoRR}, vol. abs/1908.01091, 2019.

\bibitem{shortcutlearning}
M.~Du, F.~He, N.~Zou, D.~Tao, and X.~Hu, ``Shortcut learning of large language models in natural language understanding,'' \emph{Commun.}, vol.~67, no.~1, pp. 110--120, 2024.

\end{thebibliography}
